\documentclass[prd,eqsecnum,noshowpacs,nofootinbib,preprintnumbers]{revtex4}
\usepackage{amsmath,amssymb}
\usepackage[dvips]{graphicx,color}
\usepackage{multirow}

\input epsf
\usepackage{color,graphics,latexsym,amsfonts,subfigure}

\begin{document}
\title{Bouncing cosmologies in massive gravity on de Sitter}
\author{David  Langlois, Atsushi Naruko}
\affiliation{APC (CNRS-Universit\'e Paris 7), 10 rue Alice Domon et L\'eonie Duquet, 75205 Paris Cedex 13, France;
  }

\date{\today}

\newcommand{\ma}[1]{{\mathrm{#1}}}
\newcommand{\calH}{{\cal H}}
\newcommand{\calL}{{\cal L}}
\newcommand{\calK}{{\cal K}}
\newcommand{\calR}{{\cal R}}
\newcommand{\calE}{{\cal E}}
\newcommand{\E}{{\mathbf E}}
\newcommand{\calD}{{\cal D}}
\newcommand{\pa}{{\partial}}
\newcommand{\dd}{{\rm d}}
\newcommand{\red}[1]{{\textcolor{red}{#1}}}
\newcommand{\heart}{\heartsuit}
\newcommand{\beq}{\begin{equation}}
\newcommand{\eeq}{\end{equation}}

\begin{abstract}
In the framework  of massive gravity with  a de Sitter reference metric, we study homogeneous and isotropic  solutions with positive spatial curvature. Remarkably, we find that bounces can occur  when cosmological matter satisfies the strong energy condition, in contrast to what happens in classical general relativity. This is due to the presence in the Friedmann equations of additional  terms,  which depend on the scale factor and its derivatives and can be interpreted  as  an effective fluid. 
We present a detailed study of the system using a phase space analysis. After having identified the fixed points of the system and  investigated their stability properties, we discuss the cosmological evolution in  the global physical phase space.  We find that bouncing solutions are generic. Moreover, depending on the solutions,  the cosmological evolution can lead to an asymptotic de Sitter regime, a curvature singularity or a determinant singularity. 
\end{abstract}

\pacs{04.50.Kd, 98.80.Cq}
\date{\today}
\maketitle

\section{Introduction}
After the recent discovery of a two-parameter class of massive gravity theories \cite{deRham:2010kj} that are not plagued by  the Bouldware-Deser  ghost~\cite{Boulware:1973my}, there has been an intense activity to investigate the  cosmological consequences of these models and of their extensions. In particular, it was shown that  ghost-free massive gravity with Minkowski as  fiducial metric does not admit  spatially flat FLRW solutions~\cite{D'Amico:2011jj}, although it does admit open FLRW solutions~\cite{Gumrukcuoglu:2011ew}. Using a more general fiducial metric (which also leads to ghost-free massive gravity~\cite{Hassan_etal}), one can find FLRW solutions of arbitrary spatial curvature. This applies to the interesting case of a de Sitter fiducial metric, which was studied in \cite{Fasiello:2012rw} and  \cite{Langlois:2012hk}. 
Cosmological solutions have also been investigated in other works (see e.g. \cite{cosmo_sol}). 
Beyond the question of their existence, one should stress that  the viability of  cosmological solutions  depends  on their stability properties and most cosmological models of massive gravity, including its extensions such as quasi-dilaton massive gravity~\cite{D'Amico:2012zv} and mass-varying  massive gravity~\cite{Huang:2012pe},  seem to suffer  from severe instability problems   (see e.g. \cite{DeFelice:2013bxa} and \cite{Tasinato:2013rza} for  recent reviews).  It is thus not yet clear whether a  theory of massive gravity is physically viable or not. 

In the present work, we  explore a surprising property of massive gravity. Working in the context of massive gravity defined with respect to a de Sitter metric, we show that the Friedmann equations for ordinary matter in a spacetime with positive spatial curvature  admit  {\it bouncing} solutions, which would be forbidden in classical general relativity. Indeed, let us recall that   the Friedmann equations in  Einstein gravity read 
\begin{align}
\label{friedman}
H^2=-\frac{k}{a^2}+\frac{8\pi G}{3}\rho,\qquad  \dot H+H^2=-\frac{4\pi G}{3} (\rho+3P)\,,
\end{align}
where $a(t)$ denotes the scale factor, $H\equiv \dot a/a$ the Hubble parameter and the constant $k$ characterizes the spatial curvature; $\rho$ and $P$ denote the energy density and pressure of the cosmological fluid. It is clear  that the above Friedmann equations admit bouncing solutions, i.e. a transition from a contracting phase ($H<0$) to an expanding phase ($H>0$) with a bounce characterized by  $H=0$ and $\dot H>0$,  for FLRW spacetimes with positive spatial curvature ($k>0$), but at the price of  violating   the dominant energy condition ($\rho+3P\geq 0$). 
This implies that  a bounce in classical gravity can be obtained with a scalar field in a closed FLRW spacetime (see e.g. \cite{Falciano:2008gt}). However bounces with ordinary cosmological matter (e.g. radiation or pressureless matter), which satisfies the strong energy condition, are forbidden in general relativity.  
 
 In massive gravity (on the Sitter), extra terms appear due to the presence of a potential term in the gravitational action. It is convenient to interpret these new terms in the Friedmann equations, as an effective gravitational fluid with energy density and pressure of the form~\cite{Langlois:2012hk}
  \begin{align}
  \rho_g=m_g^2\,\ {\cal E}\left(a,\frac{H}{H_c}\right) , \qquad P_g=m_g^2\ {\cal P}\left(a,\frac{H}{H_c}, \frac{\dot H}{H^2_c}\right) \,,
  \end{align}
  where $m_g$ is the mass of the graviton and $H_c$ the constant Hubble parameter characterizing the de Sitter reference metric (note that there is no explicit dependence on $a$ in the spatially flat case).   
  We stress that, despite this convenient reinterpretation,  these effective energy density and pressure do not come from an additional ad hoc exotic matter. They are purely gravitational  and fully determined  by the physical and fiducial metrics. As a consequence, the equation of state of this gravitational ``fluid'' cannot be tuned. It is thus remarkable that, for a large range of initial conditions, one obtains bouncing solutions with ordinary matter. The corresponding Friedmann equations are of the form (\ref{friedman}) with 
  \beq
  \rho=\rho_m+\rho_g,\qquad P=P_m+P_g\,,
  \eeq
  where $\rho_m$ and $P_m$ denote, respectively, the energy density and pressure of the ordinary cosmological fluid. 
 
The plan of this paper is the following. In the next section, we present our model and derive the cosmological equations of motion. In the subsequent section, we analyse in detail the evolution of the dynamical system, by identifying the fixed points and then exploring their linear stability. Depending on the values of the parameters, we draw the corresponding phase portraits. In the final section, we give a short summary and discussion of our results. 
 
\section{Setup and main equations}
We consider  a theory of massive gravity, similar to that  introduced in \cite{deRham:2010kj}, but where the reference metric is chosen to be  de Sitter instead of Minkowski. Note that de Sitter possesses as many symmetries as Minkowski spacetime, but is now characterized by a constant parameter associated with the spacetime curvature.
The gravitational action, where $g_{\mu\nu}$ denotes the usual metric minimally coupled to  matter, can be written in the form  
\begin{align}
 S_g = M_{pl}^2 \int \dd^4 x \sqrt{- g} \left[ \frac{1}{2} R
 + m_g^2 \Bigl( \calL_2 + \alpha_3 \calL_3 + \alpha_4 \calL_4 \Bigr) \right] \,,
\end{align}
 where $R$ is the  Ricci scalar for the metric $g_{\mu\nu}$ (and $g$ its determinant) and where the potential terms are given explicitly by 
 \begin{align}
 \calL_2 &= \frac{1}{2} \Bigl( [\calK]^2 - [\calK^2] \Bigr) \,, \quad
 \calL_3 = \frac{1}{6} \Bigl( [\calK]^3 - 3 [\calK] [\calK^2]
 + 2 [\calK^3] \Bigr) \,, \quad 
 \calL_4 = \frac{1}{24} \Bigl( [\calK]^4 - 6 [\calK]^2 [\calK^2]
 + 3 [\calK^2]^2 + 8 [\calK] [\calK^3] - 6 [\calK^4] \Bigr) \,,
 \end{align}
 with
 \begin{align}
 \label{K}
 \calK^\mu{}_\nu
 = \delta^\mu{}_\nu - \Bigl( \sqrt{g^{-1} f} \Bigr)^\mu{}_\nu  \,.
 \end{align}
Here the standard matrix notation is used (i.e. $(\calK^2)^\mu_{\ \nu}
 =\calK^\mu_{\ \sigma}\calK^\sigma_{\ \nu}$) and the brackets represent a trace; the square root of a matrix is defined such as 
$(\sqrt{{\cal M}})^{\, \mu}_{\ \rho} \, (\sqrt{{\cal M}})^{\, \rho}_{\ \nu} = {\cal M}^{\,\mu}_{\ \nu}$.
 
Here we focus our attention on  closed FLRW universes, whose metric can be written in the form
\begin{align}
 \dd s^2  = g_{\mu \nu} \dd x^\mu \dd x^\nu
 &= - N^2 (t) \dd t^2 + a^2 (t) \, \Omega_{i j}\,  \dd x^i \dd x^j \notag\,,
 \label{FLRW}
\end{align}
where $\Omega_{ij}$ denotes the metric of a unit 3-sphere:
\begin{align}
\Omega_{ij}\, dx^i\,  dx^j=d\chi^2+\sin^2\chi \Bigl( \dd \theta^2 + \sin^2\theta\,  \dd \phi^2 \Bigr)\,.
\end{align}

As in our previous work \cite{Langlois:2012hk},   the reference metric is taken to be de Sitter (a similar choice is  considered in \cite{Fasiello:2012rw} for flat FLRW solutions; see also \cite{Alberte:2011ah, deRham:2012kf}). In order to make the correspondence between the abstract de Sitter spacetime and the physical closed FLRW (\ref{FLRW}), we choose a coordinate system that provides a 
 closed slicing of de Sitter:
\begin{align}
\label{dS}
 \dd s^2 = f_{AB}\,  \dd X^A \dd X^B
 = - \dd T^2 + b^2 (T) \Omega_{i j} \dd X^i \dd X^j \,, \qquad \ma{where}
 \quad b (T) = \frac{1}{H_c} \cosh (H_c T) \,.
 \end{align}
 The mapping between de Sitter and our FLRW universe is described by the St\"{u}ckelberg fields $X^A=\Phi^A(x^\mu)$. With the above choice of coordinates, the FLRW symmetries are automatically satisfied by the simple prescription
\begin{align}
 \phi^0 = T = f (t) \,, \qquad \phi^i = X^i = x^i \,.
\end{align} 
Under this ansatz, the de Sitter metric is mapped into  a homogeneous and isotropic tensor in the physical spacetime, with components
\begin{align}
 f_{\mu \nu} = f_{AB}\,  \pa_\mu \phi^A \pa_\nu \phi^B 
 &= \ma{Diag} \Bigl\{ - \dot{f}^2 \,, b^2 \bigl[ f(t) \bigr] \gamma_{i j} 
 \Bigr\} \,. 
\end{align}
Inserting the two metric tensors $g_{\mu\nu}$ and $f_{\mu\nu}$ into the definition (\ref{K}), one finds that the components of  $\calK^\mu{}_\nu$ are given by
\begin{align}
 \calK^0{}_0 = 1 - \frac{\dot{f}}{N}\,, \qquad
 \calK^i{}_j = \left( 1 - \frac{b(T)}{a} \right) \delta^i{}_j \,, \qquad
 \calK^i{}_0 = 0 \,, \qquad \calK^0{}_i = 0 \,,
\end{align}
 where we have assumed here $\dot{f} > 0$.

After substitution of  the above expressions into the action, one finds that the potential part of the gravitational Lagrangian takes the  form
\begin{align}
 \calL_{mg}
 &\equiv \sqrt{- g} (\calL_2 + \alpha_3 \calL_3 + \alpha_4 \calL_4) \notag\\
 &= m_g^2 (a - b) \Bigl\{ N \Bigl[ 3 a (2 a - b) 
 + \alpha_3 (a - b) (4 a - b) + \alpha_4 (a - b)^2 \Bigr] 
 - \dot{f} \Bigl[ 3 a^2 + 3 \alpha_3 a (a - b) 
 + \alpha_4 (a - b)^2 \Bigr] \Bigr\} \,.
\end{align}
The variation of the action with respect to the time-like St\"{u}ckelberg field $f (t)$, which appears only in  $ \calL_{mg}$, thus
 yields the constraint
 \begin{align}
 \label{constraint_f}
 \left[ (3 + 3 \alpha_3 + \alpha_4)
 - 2 (1 + 2 \alpha_3 + \alpha_4) \frac{b}{a}
 + (\alpha_3 + \alpha_4) \frac{b^2}{a^2} \right] (N b' - \dot{a}) = 0 \,,
 \end{align}
 where $b$ and its derivative $b'$ are evaluated at $T=f(t)$. There are two ways to satisfy the above constraint. The first possibility is to solve algebraically the quadratic equation between the brackets. One finds in general two solutions given
by
\begin{align}
 b = X_{\pm} \, a \,, \qquad X_{\pm} = \frac{1 + 2 \alpha_3 + \alpha_4 \pm
 \sqrt{1 + \alpha_3 + \alpha_3^2 - \alpha_4}}{\alpha_3 + \alpha_4} \,,
\end{align}
which  correspond to the branches where the mass term behaves
as a cosmological constant.

 The second possibility, which we will consider in the following, is to impose that the last factor in  (\ref{constraint_f}) vanishes, i.e.
\begin{align}
 b' = \frac{\dot{a}}{N}\,.
\label{sol}
\end{align}
Let us stress that the cosmological solutions  in this branch  are necessarily accelerating, since 
  the time derivative of Eq (\ref{sol}) yields (with $N=1$)
 \begin{align}
 \ddot{a} = b'' \dot{f}= \frac{b}{H_c^2} \dot{f} > 0\,.
 \end{align}
Note that  assuming $\dot f<0$ initially  (with $ \calK^0{}_0 = 1 + \dot{f}/{N}$)  would lead to the same conclusion.
 
Using the explicit form of the function $b$, given in (\ref{dS}), one can solve  Eq (\ref{sol}) to determine the expression of  the St\"{u}ckelberg field $f(t)$:
\begin{align}
\label{f}
 f (t) = \frac{1}{H_c} \sinh^{- 1} (a H) \quad \ma{where} \quad
 H \equiv \frac{\dot{a}}{a} \,,
\end{align}
and therefore
\begin{align}
 b[f(t)] = \frac{1}{H_c} \cosh \bigl[ H_c f (t) \bigr]
 = \frac{a}{H_c} \sqrt{\frac{1}{a^2} + H^2} \,.
\end{align}
In the following, it will be convenient to introduce the new function $C$, defined as 
 \begin{align}
 C(t) \equiv \frac{1}{a(t)}b[f(t)]=\frac{1}{H_c} \sqrt{\frac{1}{a^2} + H^2} \,.
\label{def:C}
 \end{align}
By construction, we always have the inequality  $|H|/H_c\leq C$.

Variation of the total action with respect to
 the lapse function $N (t)$  and the scale factor $a (t)$ yields the following  Friedmann equations  (we have set $N=1$ {\it after} the variation):
\begin{align}  
\label{F1}
 3 H^2 + \frac{3}{a^2} &= \rho_g + \rho_m \,, \\
 \label{F2}
 - 2 \dot{H} + \frac{2}{a^2}
 &= \rho_g + P_g + \rho_m + P_m\,,
\end{align}   
 where the extra terms arising from the massive gravity potential can be interpreted as  an effective energy density and an effective pressure, defined respectively as 
\begin{align}  
\label{rho_g}
 \rho_g &\equiv - m_g^2 \left( 1 - \frac{b}{a} \right)
 \left[ 3 \left( 2 - \frac{b}{a} \right)
 + \alpha_3 \left( 1 - \frac{b}{a} \right) \left( 4 - \frac{b}{a} \right)
 + \alpha_4 \left( 1 - \frac{b}{a} \right)^2 \right] \,, \\
 \label{P_g}
   P_g
 &\equiv -\rho_g - m_g^2 \left(\dot{f} - \frac{b}{a} \right)
 \left[ 3 - 2 \frac{b}{a} 
 + \alpha_3 \left( 1 - \frac{b}{a} \right) \left( 3 - \frac{b}{a} \right)
 + \alpha_4 \left( 1 - \frac{b}{a} \right)^2 \right] \,.
\end{align}   
One can obtain expressions that depend explicitly on the scale factor and its derivatives by replacing the ratio $b/a$ with $C$ and $\dot f$ by 
\begin{align}
\dot f= \frac{a(\dot H+H^2)}{H_c\sqrt{1 +a^2H^2}}\,,
\end{align}
which follows from Eq.~(\ref{f}).

In the following we will restrict ourselves to   the case $\alpha_3 = \alpha_4 = 0$  for simplicity.
The evolution equations (\ref{F1})-(\ref{F2}), together with (\ref{rho_g})-(\ref{P_g}),  can be rewritten in terms of $C$ and $H$ as
\begin{align}  
 &  3 \Bigl[ (1 + \lambda) C^2 - 3 \lambda C
 + 2 \lambda \Bigr]
 = \calR_m\,,
  \\
 & \frac{1}{H_c^2} \frac{\ddot{a}}{a}=\frac{\dot{H}}{H_c^2}+ \frac{H^2}{H_c^2} =- \frac{C}{C - C_H} \left[ 
 \frac{1 + 3 w}{2}  C^2 
 - (2 + 3 w) C_H C + 2 (1 + w) C_H \right]\,,
\label{eq0:H}   
\end{align}
 where we have introduced the dimensionless quantities 
 \begin{align}
 \calR_m \equiv \frac{\rho_m}{H_c^2} \,, \qquad
  w \equiv \frac{P_m}{\rho_m} \,, \qquad 
 \lambda \equiv \frac{m_g^2}{H_c^2} \,, \qquad
 C_H \equiv \frac{3 \lambda}{2 (1 + \lambda)}\,.
 \end{align}
 The variable $\calR_m$ is time dependent and its evolution is determined by the energy conservation equation, which reads
 \begin{align}  
 \dot \calR_m+3(1+w)  H \calR_m=0\,,
 \end{align}  
 and we will assume an equation of state that satisfies the strong energy condition,  i.e. $-1/3< w \leq 1$.
 
 Using the relation 
 \begin{align}
 \label{Cdot}
 \dot{C} = \frac{H}{C} \left( \frac{\dot{H}}{H_c^2}
 + \frac{H^2}{H_c^2} - C^2 \right) \,,
\end{align}
which follows from the definition Eq (\ref{def:C}), one can write down an equation of motion for $C$ in the form
\begin{align}
 \dot{C}
 &= - \frac{(1 +  w) H}{C - C_H} \left[  \frac{3}{2}  C^2 
 - 3 C_H C + 2 C_H \right]\,.
\end{align}
Note that the fact that our cosmological solutions must be accelerating, i.e. $\ddot a>0$, as discussed earlier, means that the right hand side of (\ref{eq0:H}) is restricted to be positive.

\section{Analysis of the dynamical system}

We now investigate in a systematic way the dynamical evolution of the system introduced in the previous section. We would like to know, starting from arbitrary initial conditions, how the Universe is going to evolve and what  its final  fate will be. 
Since the evolution of the system is determined by the two time-dependent functions $H (t)$ and $C (t)$, it is convenient to describe the cosmological evolution  in a two-dimensional diagram. To understand the generic behaviour of this dynamical system, it is crucial to identify its fixed points, which we do just below. In the second part of this section, we study the behaviour of the system in the vicinity of  these fixed points and then discuss qualitatively the global cosmological evolution by plotting phase portraits.

Let us first introduce a dimensionless Hubble  parameter and a dimensionless time, defined respectively by
\beq
h\equiv \frac{H}{H_c}\,, \qquad \tau\equiv H_c\, t\,.
\eeq
The cosmological evolution can  be rewritten in terms of dimensionless quantities as follows:
\begin{align}  
 \label{eq:C}
 \frac{\dd C}{\dd \tau}
 &= f_C(h,C)\equiv  - \frac{(1 +  w)\,  h}{C - C_H}\left[  \frac{3}{2}  C^2 
 - 3  C_H C + 2  C_H \right]
 \\
 \label{eq:H}
 \frac{\dd h}{\dd \tau}
 &=f_h(h,C)\equiv  - h^2 -  \frac{C}{C - C_H} \left[ 
 \frac{1 + 3 w}{2}  C^2 
 - (2 + 3 w) C_H C + 2 (1 + w) C_H \right]\,.
\end{align}
Note that this system is invariant under the changes  $h\rightarrow -h$ and  $\tau \rightarrow -\tau$, which means that the regions $h>0$ and $h<0$ of the phase diagram are symmetric up to a time reversal.

\subsection{Fixed points}
\begin{center}
\begin{figure}
\includegraphics[width=10cm]{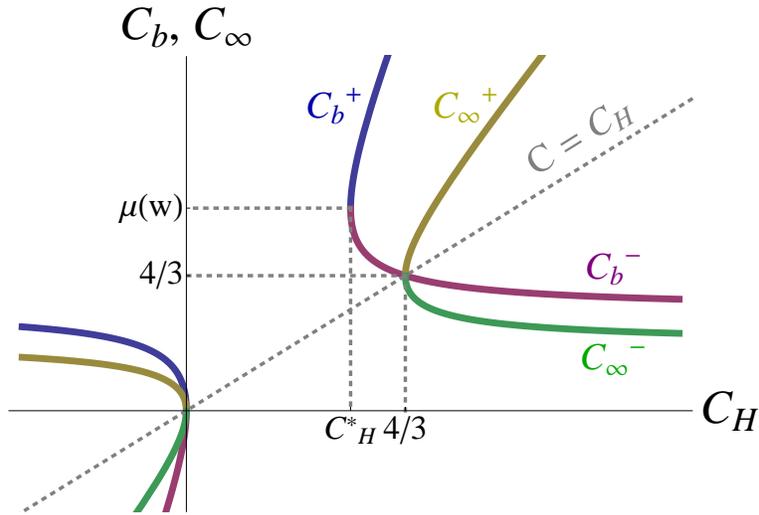}
\caption{The values of the roots $C_b^\pm$ and  $C_\infty^\pm$ are plotted as functions of $C_H$. The plot corresponds to the value $w=0$ but the relative position of these four quantities remain unchanged for other values of $w$ in the range $0\leq w \leq 1$.
}
\label{fig:cbcinf}
\end{figure}
\end{center}

Let us first identify  the fixed points satisfying  $h=0$, in which case the right hand side of (\ref{eq:C}) vanishes and 
the right hand side of  Eq (\ref{eq:H}) reduces to
\begin{align}  
 f_h(h=0,C)= -  \frac{C}{C - C_H} \left[ 
 \frac{1 + 3 w}{2}  C^2 
 - (2 + 3 w) C_H C + 2 (1 + w) C_H \right]\,.
\end{align}
The fixed points correspond to the values of $C$ (together with $h=0$) for which the above expression vanishes. We thus find $C=0$ and the solutions of the quadratic equation inside the brackets,
\begin{align}
 C_b^{\pm} = \frac {(2 + 3 w) C_H \pm \sqrt{(2 + 3 w)^2 C_H^2
 - 4 (1 + 3 w) (1 + w)C_H}}{(1 + 3 w)} \,,
\end{align}
which  exist  if $C_H\leq 0$ or if 
\begin{align}
C_H\geq  C_H^*(w),\qquad C_H^*(w) \equiv \frac{4 (1 + 3 w) (1 + w)}{(2 + 3 w)^2} < 4/3\,.
\end{align}
Note that when $C_H=C_H^*(w)$, or equivalently $\lambda=\lambda^*(w)$ with 
\beq
\lambda^*(w)\equiv\frac{8(1+w)(1+3w)}{4+4w+3w^2}\,,
\eeq
the two solutions $C_b^+$ and $C_b^-$ coincide and are given by
\beq
C_b^\pm=\mu(w)\equiv \frac{4(1+w)}{2+3w} \qquad {\rm for} \quad  C_H=C_H^*(w)\,.
\eeq
 The dependence of $C_b^\pm$ on $C_H$, or equivalently on $\lambda$, is illustrated in Fig. \!\ref{fig:cbcinf}.

It is also possible to find fixed points away from the axis $h=0$. Eq~(\ref{Cdot}) implies that such fixed points must satisfy $C^2=H^2/H_c^2$. Note that this condition corresponds, according to the definition (\ref{def:C}) of $C$,  to the  limit of $C$ when $a\gg H^{-1}$.
Substituting this condition into the right hand side of Eq (\ref{eq:C}) and imposing that it vanishes leads to the equation
\begin{align}
3 C^2 - 6 C_H C + 4 C_H = 0 \,,
\label{eq:bounce}
\end{align}
with  solutions\footnote{We use the subscript $\infty$ to keep in mind that they correspond to the asymptotic limit $a\gg 1/H$.}
\begin{align}
 C_\infty^\pm = C_H \pm \sqrt{C_H \left( C_H - \frac{4}{3} \right)} =\frac{3\lambda \pm \sqrt{\lambda (\lambda-8)}}{2(1+\lambda)} \,.
\label{def:Cinf}
 \end{align}
 These solutions exist provided 
 \begin{align}
C_H\leq 0, \qquad C_H\geq 4/3,
 \end{align}
 which correspond, respectively, to the intervals $-1<\lambda \leq 0$ and  $\lambda <-1$ or $\lambda\geq 8$. The dependence of $C_\infty^\pm$ on $C_H$ is also plotted in Fig. \!\ref{fig:cbcinf}.

Before closing this subsection, it is instructive
 to rewrite the basic equations in terms of $C_b^\pm$ and $C_\infty^\pm$,
 assuming their existence. The equations of motion read
 \begin{align}
 \frac{\dd C}{\dd \tau}
 &= f_C (h, C)
 = - \frac{3 (1 +  w)}{2} \frac{h}{C - C_H} (C - C_\infty^+) (C - C_\infty^-)\,,
  \\
 \label{fh2}  
 \frac{\dd h}{\dd \tau}
 &= f_h (h, C)
 = - h^2 - \frac{1 + 3 w}{2} \frac{C}{C - C_H} (C - C_b^+) (C - C_b^-)\,,
\end{align}
while  the  two constraints are now given by 
 \begin{align}  
 \label{constraint_M}
 \calR_m
 &=\frac{9}{3-2C_H} (C - C_\infty^+) (C - C_\infty^-) \geq 0 \qquad (M)\,,
 \\
  \label{constraint_A}
 \frac{1}{a} \frac{\dd^2 a}{\dd \tau^2}
 &= - \frac{1 + 3 w}{2} \frac{C}{C - C_H} (C - C_b^+) (C - C_b^-) \geq 0 \qquad (A)\,.
\end{align}
 
 Before investigating the stability of the fixed points in the next subsection, let us mention another special value of $C$, namely $C_H$, where the equations of motion become singular. The limit $C\rightarrow C_H$ corresponds to a curvature singularity, since $\dot H$ diverges in this limit. We encountered the same curvature singularity in our study of flat FLRW solutions in massive gravity on de Sitter~\cite{Langlois:2012hk}.

\subsection{Stability  of the fixed points}
In order to understand qualitatively the evolution in the phase space, it is useful to  study the evolution  of linear perturbations  around the fixed  points that we have identified. 
Introducing the perturbations
 \begin{align}
 \delta C \equiv C - C_\ma{fixed} \,, \qquad  \delta h \equiv h - h_\ma{fixed}\,,
 \end{align}
 and linearizing the evolution  equations
 with respect to $\delta C$ and $\delta h$,
 we obtain a linear system of the form
 \begin{align}
 \frac{\dd}{\dd \tau}
 \begin{pmatrix} 
 \delta C \\
 \delta h
 \end{pmatrix} 
 = \begin{pmatrix} 
 \displaystyle \frac{\pa f_C}{\pa C} &
 \displaystyle \frac{\pa f_C}{\pa h} \\
 \displaystyle \frac{\pa f_h}{\pa C} &
 \displaystyle \frac{\pa f_h}{\pa h} 
 \end{pmatrix}_\ma{fixed}
 \begin{pmatrix} 
 \delta C \\
 \delta h
 \end{pmatrix} 
 \quad 
 \equiv \quad \mathbf{E}\ 
\begin{pmatrix} 
 \delta C \\
 \delta h
 \end{pmatrix} \,,
\end{align}
where $\mathbf{E}$ is a short notation for the $2 \times 2$ matrix on the right hand side.
The stability properties of the system around any fixed point can be deduced from the eigenvalues of its evolution matrix $\E$, 
\begin{align}
 {\cal E}_{1,2}= \frac 12\left( {\mathrm Tr}(\E)\pm \sqrt{{\mathrm Tr}(\E)^2-4 {\mathrm Det}(\E)}\right)\,.
 \end{align}
If ${\mathrm Det}(\E)<0$, the eigenvalues are real and have different sign: the fixed point is a saddle point. If ${\mathrm Det}(\E)>0$, the two eigenvalues, when real,  have the same sign, which is given by ${\mathrm Tr}(\E)$: if  both eigenvalues are negative, the fixed point is an attractor because all flows   converge towards it; conversely,  if both  eigenvalues are positive, the fixed point is a repeller. 
Moreover, depending whether the eigenvalues are real or imaginary, the fixed point is said to be nodal or spiral, respectively.

Let us first consider the fixed points defined by  $C=C_b^\pm$, which are located on the axis $h=0$. 
Their  evolution matrix reads 
\begin{align}
\E
 = 
 \begin{pmatrix}
 0 & \displaystyle \frac{\pa f_C}{\pa h} \\
 \displaystyle \frac{\pa f_h}{\pa C} & 0 \\
 \end{pmatrix} 
_{(C_b^\pm, 0)}
\end{align}
 with
\begin{align}
 \frac{\pa f_C}{\pa h}
 &= - \frac{3 (1 + w)}{1 + 3 w} \frac{C_H}{C_b^\pm - C_H}
 \left( C_b^\pm - \frac{4}{3} \right) \,, \qquad
 \frac{\pa f_h}{\pa C}
 = - \frac{ (2 + 3 w) C_H}{C_b^\pm - C_H}
 \left[ C_b^\pm - \mu(w) \right]\,.
\end{align}
The trace and the determinant of the matrix are thus 
\begin{align}
{\mathrm Tr}(\E)=0\,,\qquad  {\mathrm Det}(\E)= -\frac{\pa f_C}{\pa h} \frac{\pa f_h}{\pa C}=- \frac{3 (1 + w) (2 + 3 w)}{1 + 3 w} \frac{C_H^2}{(C_b^\pm - C_H)^2}
 \left( C_b^\pm - \frac{4}{3} \right)
 \left(C_b^\pm - \mu(w) \right)\,.
\end{align}
The eigenvalues are real if ${\mathrm Det}(\E)<0$, giving a saddle point; purely imaginary if ${\mathrm Det}(\E)>0$, in which case the fixed point is spiral. According to Fig.\ref{fig:cbcinf}, one can easily see that the case ${\mathrm Det}(\E)>0$ is possible only  for the fixed point $C_b^-$, provided $C_H$ is the range $C_H^*(w)< C_H< 4/3$. In all other cases, the fixed points $C_b^\pm$ are saddle points.

One can proceed in a similar manner for the fixed points $C_\infty^\pm$. Their evolution matrix reduces to 
\begin{align}
\E
 = 
 \begin{pmatrix}
 \displaystyle \frac{\pa f_C}{\pa C} & 0 \\
 \displaystyle \frac{\pa f_h}{\pa C} &
 \displaystyle \frac{\pa f_h}{\pa h} \\
 \end{pmatrix} _{(C_\infty, h_\infty)}
\end{align}
 with
\begin{align}
 \frac{\pa f_C}{\pa C}
 &= - 3 (1 + w) h_\infty \,, \qquad
 \frac{\pa f_h}{\pa C}
 = - (1 + 3 w) C_\infty \,, \qquad
 \frac{\pa f_h}{\pa h} 
 = -2h_\infty\,,
\end{align}
 where $h_\infty=\pm C_\infty$.
The two eigenvalues of the matrix are  simply $ - 3 (1 + w) h_\infty$ and $ - 2 h_\infty$. The fixed points in the region $H>0$ are thus attractors while the fixed points in the $H<0$ are repellers. 

Let us mention briefly the fixed point at the origin, which is a saddle point. Since it is located at a corner of the physical phase space, only the quadrant on its right is relevant.

It is also worth noticing that the vector field tangent to the flow becomes vertical on the axes $C=C_\infty^\pm$, which means that trajectory cannot cross these axes. As a consequence, if the constraint (\ref{constraint_M}) is satisfied initially, it will remain so in the future. However, this is not the case for the constraint (\ref{constraint_A}). Indeed, one can check that the flow is not vertical on the axes $C=C_b^\pm$, so that a trajectory can a priori cross these boundaries in phase space. Therefore, even if the constraint $\ddot a>0$ is satisfied initially, some of the phase space trajectories  can reach $\ddot a=0$ during their evolution. Although this transition is perfectly regular from the point of view of Friedmann's equations,  it is problematic from the point of view of the underlying massive gravity theory where $\ddot a=0$ corresponds to $\dot f=0$, in which case the tensor $f_{\mu\nu}$ becomes degenerate. In other words, the coordinate transformation between the abstract de Sitter space and the FLRW coordinates, embodied by the St\"uckelberg fields, becomes singular: $\det(\partial_\mu\phi^A)=0$.  It is intriguing that this determinant singularity, following the terminology used in \cite{Gratia:2013gka},  which signifies a breakdown of the massive gravity framework, at least in its present formulation,  does not appear at all in the cosmological evolution at the level of the Friedmann equations. 

\subsection{Global analysis}
After having analysed the stability of the various fixed points separately, we  now present a global discussion on the existence of the fixed points and their stability,  depending on the value of the parameter $C_H$ (or equivalently $\lambda$). In each case, we will draw the corresponding phase portrait.

As discussed earlier, and illustrated in Fig. \!\ref{fig:cbcinf}, the existence of the various fixed points that we have identified depends on the value of $C_H$. The fixed points $C_b^\pm$ exist only if
$C_H\leq 0$ or 
$C_H\geq  C_H^*(w)$, whereas  the existence of the fixed points $C_\infty^\pm$  requires $C_H\leq 0$ or  $C_H>4/3$. Taking also  into account the special value $C_H=3/2$ (corresponding  to infinite  $\lambda$) for which the constraint (M) is not defined, one can  distinguish five intervals for the parameter $C_H$, separated by the values $3/2$, $4/3$, $C_H^*(w)$ and  $0$, in decreasing order. 

\subsubsection{Case $C_H>3/2$ (or $\lambda< -1$)}

\begin{figure}
\subfigure[Global picture]{%
\includegraphics[clip, width=8cm]{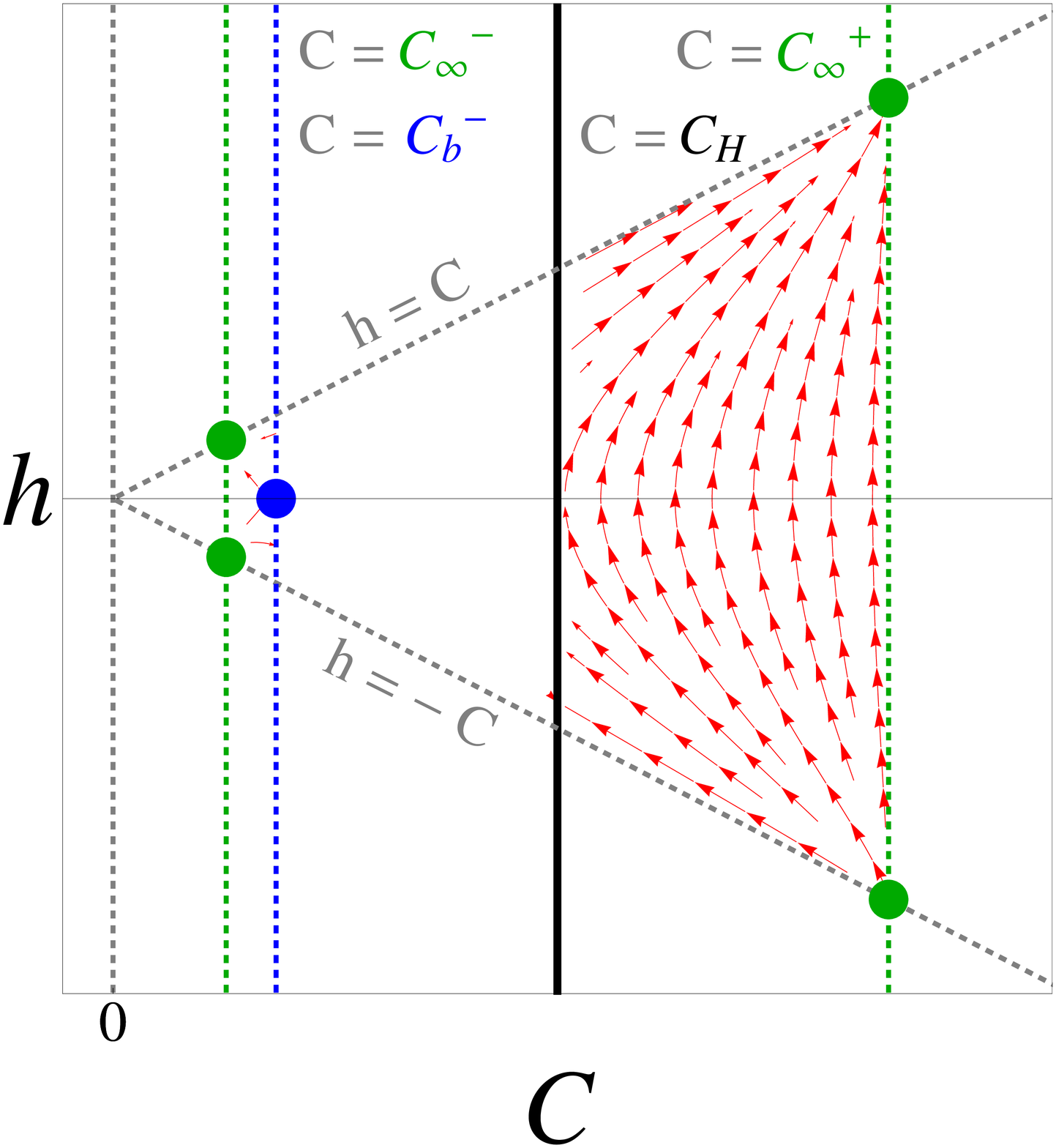}}%
\qquad \qquad 
\subfigure[Enlargement of the region $C_\infty^- \leq C \leq C_b^-$]{%
\includegraphics[clip, width=8cm]{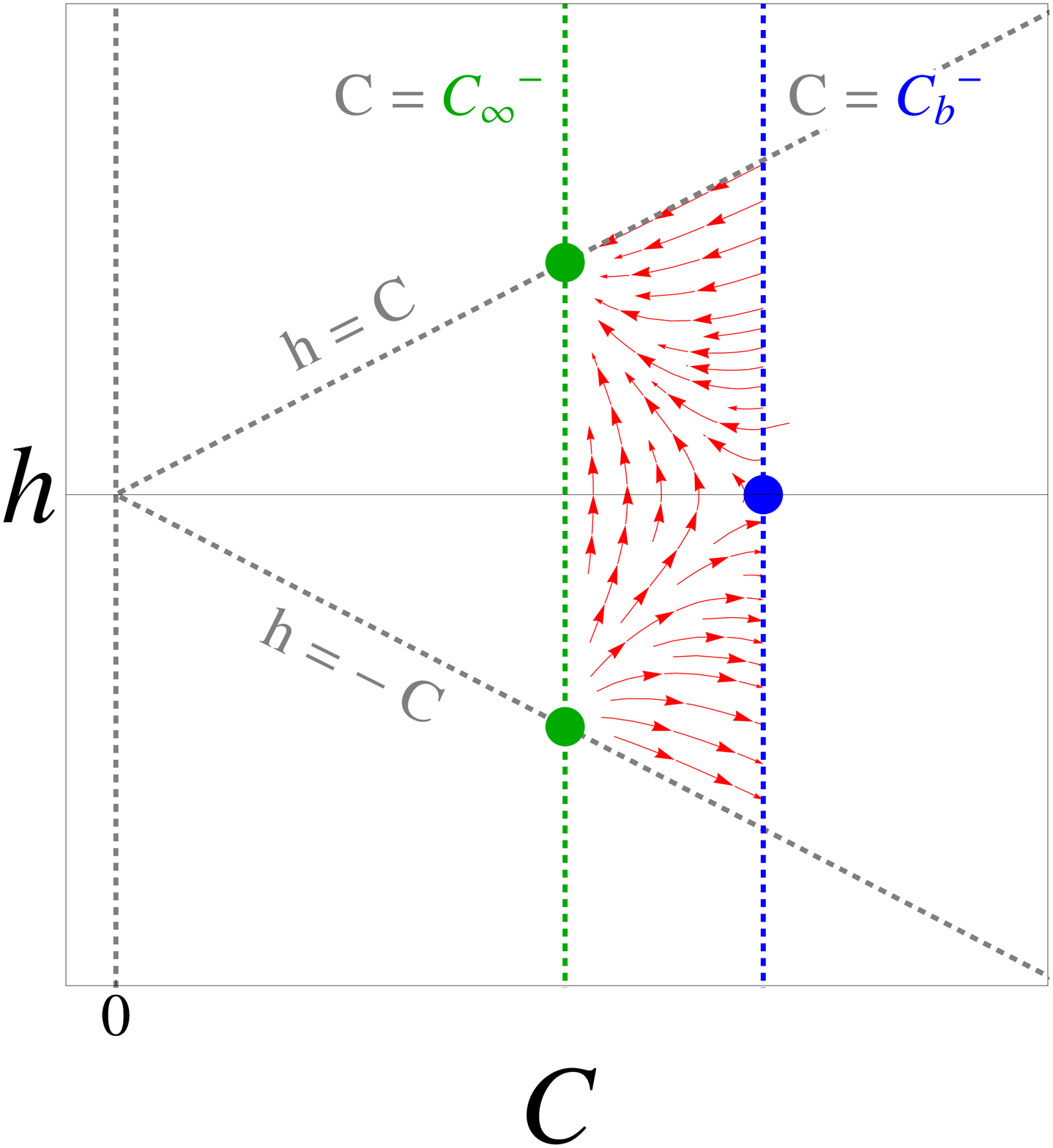}}%
\caption{Phase diagram in the case $C_H > 3/2$ (i.e.  $\lambda< -1$). All fixed points are defined, but the fixed point $(C_b^+,0)$ is outside of the region allowed by the constraint (M), and is therefore not represented.  The region between $C_b^-$ and $C_H$ is forbidden by the constraint (A). The trajectories that approach $C=C_H$ are singular in this limit, which means they cannot cross the boundary $C=C_H$.}
\label{fig:case1}
\end{figure}

This case is illustrated in Fig. \ref{fig:case1}.
All fixed points $C_b^\pm$ and $C_\infty^\pm$ exist and  they are ordered as 
\begin{align}
\label{order1}
0<C_\infty^- \leq C_b^- \leq C_H \leq C_\infty^+ < C_b^+\,.
\end{align}
The second constraint (M) implies that the trajectory must be inside the region 
\begin{align}
C_\infty^- \leq C \leq C_\infty^+\qquad (M)
\end{align}
which contains $C_b^-$ but not $C_b^+$. 
Within this region, the system accelerates, i.e. satisfies the constraint (A), only in the two subregions
\begin{align}
 C_\infty^- \leq C \leq C_b^- \quad \ma{and} \quad C_H \leq C \leq C_\infty^+  \qquad (M) \, \& \, (A)\,.
\end{align}

In the first region, one finds three fixed points: the repeller $(C_\infty^-, -C_\infty^-)$, the attractor $(C_\infty^-, +C_\infty^-)$ and $(C_b^-,0)$ on the axis. According to our discussion of the previous subsection, the eigenvalues for the fixed point $C_b^-$ are real  (and opposite) because $ {\mathrm Det}(\E)<0$. The cosmological flows in this region are depicted in Fig. \ref{fig:case1}. As one can see, there exist some solutions that reach the determinant singularity on the axis $C=C_b^-$. 

In the region $C_H \leq C \leq C_\infty^+$, one  finds trajectories that start in the asymptotic past at the fixed point $(C_\infty^-, -C_\infty^-)$, in a contracting phase, and evolve to reach in the asymptotic future  the symmetric fixed point $(C_\infty^-, C_\infty^-)$. They represent perfect examples of  bouncing cosmologies with a regular behaviour. Other trajectories end, or start, on the boundary $C=C_H$, which corresponds to a curvature singularity (where $\dot H$ diverges) and  do not experience bouncing.

\subsubsection{Case $4/3\leq C_H<3/2$ (or $\lambda\geq 8$)}
In this parameter range,  we still have the existence of all fixed points with the same ordering (\ref{order1}) as in the previous case. However, the constraint (M) now {\it excludes} the region between $C_\infty^-$ and $C_\infty^+$, so that the allowed region is 
\begin{align}
C\leq C_\infty^-\quad {\rm or }\quad C\geq C_\infty^+ \qquad (M)\,.
\end{align}
The fixed point $C_b^-$  and the singular line $C=C_H$ are  thus irrelevant in this case. Taking into account the constraint (A) as well, the second region gets restricted so that the final allowed regions are
\begin{align}
  C \leq C_\infty^- \quad \ma{and} \quad C_\infty^+ \leq C \leq C_b^+ \qquad
 (M) \, \& \, (A)\,.
\end{align}

\begin{center}
\begin{figure}
\includegraphics[width=8cm]{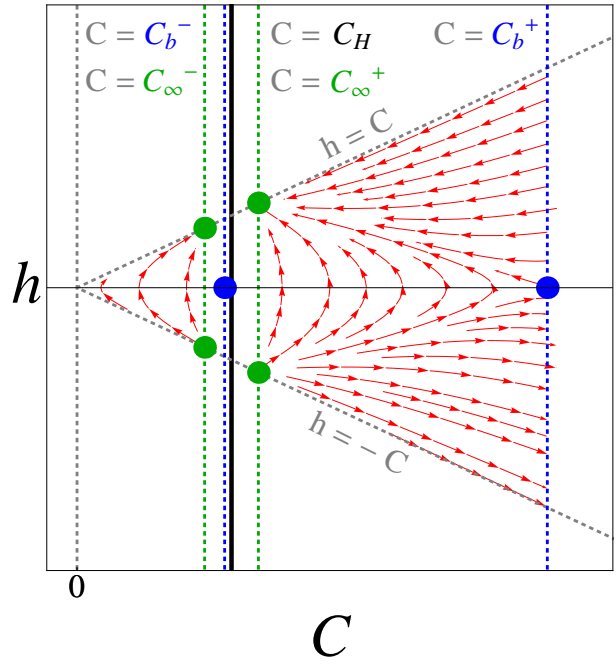}
\caption{Phase diagram in the case $4/3<C_H <3/2$ (i.e.  $\lambda> 8$). All fixed points are defined. The region between $C_\infty^-$ and $C_\infty^+$, which contains the fixed point $(C_b^-,0)$ and the axis $C=C_H$,  is forbidden by the constraint (M).  The region above $C_b^+$  is forbidden by the constraint (A). In the left region, all solutions are bouncing and relate  $(C_\infty^-,-C_\infty^-)$ to $(C_\infty^-,C_\infty^-)$. In the right region, one finds bouncing solutions that evolve from $(C_\infty^+,-C_\infty^+)$ to $(C_\infty^+, C_\infty^+)$, but also solutions that evolve into or from the 
determinant singularity.
}
\label{fig:case2}
\end{figure}
\end{center}

In the region $C \leq C_\infty^-$,  all trajectories are bouncing solutions, which start from the repeller $(C,h)=(C_\infty^-, -C_\infty^-)$ and end at the attractor $(C,h)=(C_\infty^-, C_\infty^-)$.  In the second region, $C_\infty^+ \leq C \leq C_b^+$, there exist  bouncing trajectories between the fixed points $(C_\infty^+, -C_\infty^+)$ and $(C_\infty^+, C_\infty^+)$, but also trajectories that reach the boundary $C_b^+$. For the fixed  point $C_b^+$, we find that the eigenvalues are  real (and opposite) as in the previous case. The phase diagram  is depicted in Fig. \ref{fig:case2}.

We should consider separately the degenerate case $C_H=4/3$, for which
\begin{align}
C_\infty^+=C_\infty^-=C_b^-=C_H=\frac43\,.
\end{align}
The two repellers ($h<0$) merge into a single repeller at $(4/3, -4/3)$. Similarly, the two attractors with $h>0$ merge into a single attractor at $(4/3, 4/3)$. Moreover, the fixed point $C_b^-$ disappears since the right hand side of (\ref{fh2}) reduces to 
\beq
f_h (h, C)
 = - h^2 - \frac{1 + 3 w}{2} C (C - C_b^+)\,.
 \eeq

\begin{center}
\begin{figure}
\includegraphics[width=8cm]{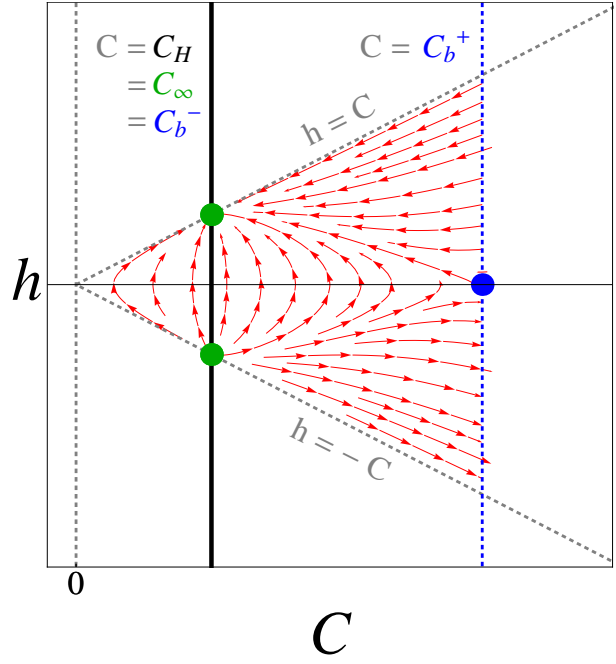}
\caption{Phase diagram in the special case $C_H =4/3$ (i.e.  $\lambda=8$). The fixed  points $C_\infty^-$ and $C_\infty^+$ now coincide, so that the intermediate forbidden region of  Fig. \ref{fig:case2} has  disappeared. Apart this difference, the behavior in the rest of the phase space is quite similar to that of Fig. \ref{fig:case2}.}

\label{fig:case3}
\end{figure}
\end{center}

The exclusion region due to the constraint (M) disappears and the allowed region satisfying the constraint (A) is simply 
\begin{align}
 0 < C \leq C_b^+ \qquad (A)\,.
\end{align}
The qualitative behaviour is exactly the same as in the allowed region of the non degenerate case, as illustrated in 
 Fig. \ref{fig:case3}.

\subsubsection{$C_H^*(w)\leq C_H< 4/3$ (i.e. $\lambda^*(w) \leq \lambda <8$)}
In this case, the fixed points $C_\infty^\pm$ do not exist and we are left with  the fixed points $C_b^\pm$ on the $H=0$ axis. The hierarchy is now given by
\begin{align}
C_H<C_b^-\leq C_b^+\,.
\end{align}
Since $C_H<3/2$ and the two roots $C_\infty^+$ and $C_\infty^-$ are imaginary, the constraint (M) is  automatically satisfied, whereas  the constraint (A) implies 
\begin{align}
 C \leq C_H \quad \ma{or} \quad C_b^- \leq C \leq C_b^+ \qquad (A)\,.
\end{align}
We thus find an excluded region between $C_H$ and $C_b^-$. 

\begin{center}
\begin{figure}
\includegraphics[width=8cm]{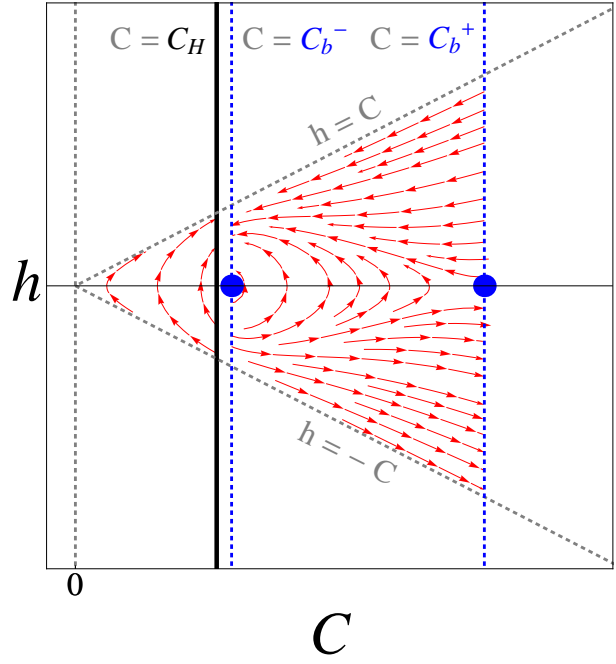}
\caption{Phase diagram in the case $C_H^*(w)\leq C_H< 4/3$ (i.e. $\lambda^*(w) \leq \lambda <8$).  The fixed points $C_\infty^\pm$ do not exist.  Two regions are forbidden by the constraint (A): the first one  between $C_H$ and $C_b^-$, the second one beyond $C_b^+$. All solutions start and end on singular boundaries: curvature singularity in the left region and determinant singularity in the right region.
}

\label{fig:case4}
\end{figure}
\end{center}

The behavior of the perturbations around  $C_b^-$  also differs from the previous cases since we now have $4/3<C_b^-< \mu(w)$, which implies  ${\mathrm Det}(\E)>0$. 
The eigenvalues for $C_b^-$ are thus purely imaginary, which corresponds to a spiral point, as illustrated 
 in Fig. \ref{fig:case4}. The other fixed point $C_b^+$ remains a saddle point. 
In the region $C \leq C_H$, there is no fixed point (except the origin) and the trajectories evolve from the line $C=C_H$ with $h<0$ towards the same line with $h>0$. All  solutions correspond to bouncing solutions but they are singular in the past and in the future. In the second region, $C_b^- \leq C \leq C_b^+$, one also finds bouncing trajectories starting and ending on the line $C=C_b^-$. We also find solutions that are not bouncing: contracting solutions that  go from the line $C=C_b^-$ to the line $C=C_b^+$ and expanding solutions  that  go from  $C=C_b^+$ to the boundary $C=C_b^-$.

\begin{center}
\begin{figure}
\includegraphics[width=8cm]{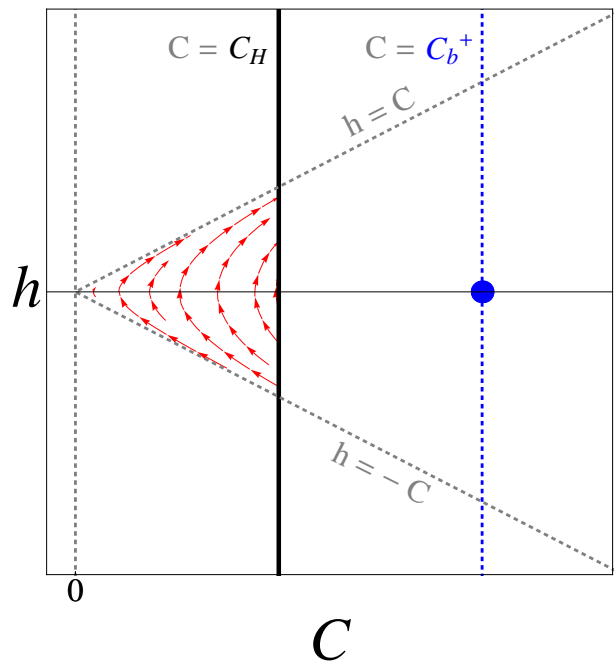}
\caption{Phase diagram in the special case $C_H =C_H^*(w)$ (i.e.  $\lambda=\lambda^*(w)$). The fixed  points $C_b^-$ and $C_b^+$ now coincide, but are in the region forbidden by (A). All solutions in the allowed region are bouncing, evolving from  the negative part to the positive part of the $C=C_H$ axis, characterized by a curvature singularity.}
\label{fig:case5}
\end{figure}
\end{center}

Finally, let us discuss the  special case where the two fixed points $C_b^-$ and $C_b^+$ become degenerate, which occurs for the value 
$C_H=C_H^*(w)$, in which case 
\begin{align}
C_b^\pm=\mu(w)\,.
\end{align}
The  constraint (A) then imposes
\begin{align}
 C < C_H \,,
\end{align}
which means that the only fixed point $C_b^-\equiv C_b^+$ is outside the allowed region.
As illustrated in   Fig. \ref{fig:case5}, the situation is quite analogous to what happens in the left region ($C < C_H$) of the non degenerate case.

\subsubsection{$0< C_H<C_H^*(w)$ (i.e. $0< \lambda< \lambda^*(w)$)}
There is no fixed point in this case and  the constraint (A) imposes the condition
\begin{align}
 C \leq C_H \,. 
\end{align}
The evolution is phase space is similar to the special limit  $C_H=C_H^*(w)$ discussed previously. Once again, all solutions are bouncing and connect the lower $C = C_H$ axis  to its upper part.

\subsubsection{$C_H<0$ (i.e. $-1<\lambda<0$)}
\begin{center}
\begin{figure}
\includegraphics[width=8cm]{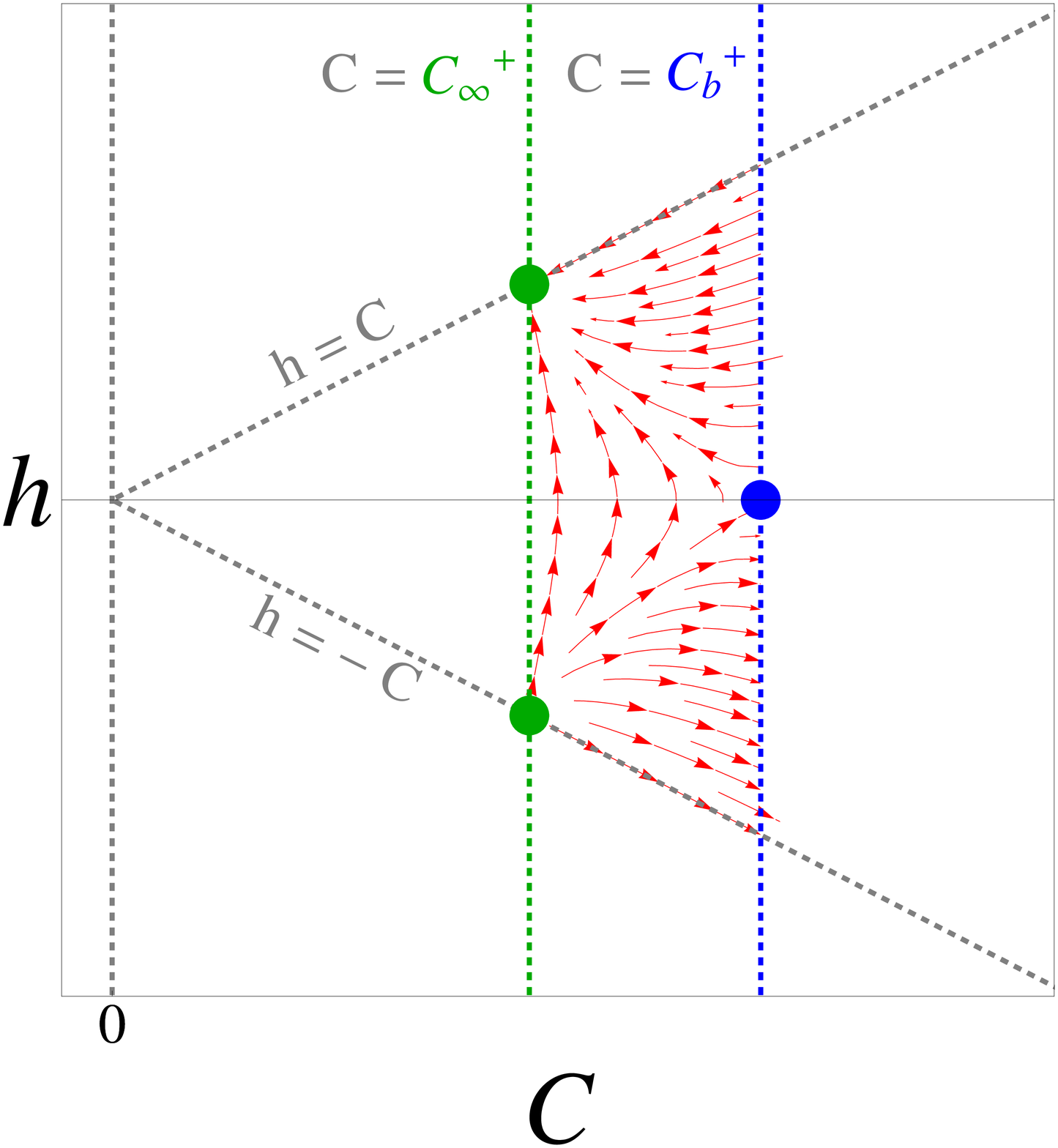}
\caption{Phase diagram in the case $C_H<0 $ (i.e.  $-1<\lambda<0$). There are three fixed points and the phase portrait is quite similar to that of the right region in Fig.\ref{fig:case2}.}
\label{fig:case6}
\end{figure}
\end{center}

In this parameter range, only the fixed points $C_\infty^+$ and $C_b^+$ are defined, since the solutions $C_\infty^-$ and $C_b^-$ are negative. 
Consequently, the constraint (M) imposes the restriction
\beq
C\geq C_\infty^+\qquad (M)
\eeq
while the constraint (A) requires 
\beq
C< C_b^+\qquad (A)\,.
\eeq
The allowed region is therefore between $C_\infty^+$ and $C_b^+$. As illustrated in Fig. \ref{fig:case6}, one finds bouncing solutions evolving from 
the fixed point $(C_\infty^+, -C_\infty^+)$ towards $(C_\infty^+, C_\infty^+)$. The other solutions start or end at the boundary $C=C_b^+$ corresponding to the determinant singularity. 

\section{Summary and discussion}
In this paper, we have studied closed FLRW solutions in the context of massive gravity with de Sitter as fiducial metric. We have found that bouncing solutions are generic when  cosmological matter satisfies the strong energy condition ($w>-1/3$). This is in contrast with classical general relativity where closed bouncing solutions  require cosmological matter that violates the strong energy condition. The different outcome  in the  massive gravity framework is due to  the presence of extra terms, which arise from the potential term in the gravitational action. 

We have also investigated the dynamical evolution of the cosmological system by resorting to  standard phase space analysis. Introducing a  two-dimensional phase space spanned by the variables $C$ and $h=H/H_c$, we have identified the fixed points of the system. Their number depends on $\lambda = m_g^2/H_c^2$, i.e. the square of the ratio between the graviton mass and the fiducial Hubble parameter. The maximal number of fixed points (ignoring the origin) is six: two on the axis $h=0$ ($C_b^-$ and $C_b^+$), and two pairs of  points symmetric with respect to the $h=0$ axis and localized on the $h=\pm C$ lines. There also exists a vertical line $C=C_H$ where the solutions encounter a curvature singularity. 
For all values of the parameter $\lambda$, we have found that bouncing solutions are generic. 

In summary, the cosmological solutions can have three different fates (and corresponding origins, by time reversal). These fates are analogous to those encountered in the spatially flat case, discussed in our previous work~\cite{Langlois:2012hk}. The first fate is a regular asymptotic de Sitter regime (corresponding to $C_\infty$). The second possibility is that the Universe ends in a curvature singularity ($C=C_H$). Finally, the third possibility is to reach a singularity specific to massive gravity, on the boundary $C=C_b$,  where  $\dot f=0$ and the tensor $f_{\mu\nu}=0$ thus becomes degenerate. Intriguingly, there is no trace of this singular behaviour at the level of the effective Friedmann equations. This means that if one works directly with the generalized Friedmann equations, without worrying about the underlying theory, one can simply continue the cosmological evolution across this boundary where nothing seems to happen, except that the solution is decelerating after the transition\footnote{Mathematically, this would mean that 
one can still apply the definition of ${\cal K}^0_0$ in Eq.~(\ref{K}) even if $\dot f<0$, in contradiction with the usual prescription for the square root.}.
 This might indicate that this type of singularity  could be resolved in an  extended framework of massive gravity, as suggested recently in \cite{Gratia:2013gka}.

\acknowledgements
A.N. was supported by JSPS Postdoctoral Fellowships for Research Abroad. 
D.L. was partly supported by the ANR (Agence Nationale de la Recherche)
 grant STR-COSMO ANR-09-BLAN-0157-01.
The authors thank the Yukawa Institute for Theoretical Physics, Kyoto University, for hospitality, in particular during  the YITP workshop YITP-T-12-04 on
 ``Nonlinear massive gravity theory and its observational test''
 and YITP-T-12-03 on ``Gravity and Cosmology 2012''.

\end{document}